\documentclass[aps,pre, preprint,superscriptaddress]{revtex4-1}

\usepackage{graphicx, amssymb, verbatim, amsmath, siunitx, color, float, soul, makecell, xcolor}
\usepackage{natbib,url, enumerate}
\usepackage[colorlinks, allcolors = blue, bookmarksopen, bookmarksnumbered]{hyperref}

\definecolor{colsq}{rgb}{0,0.4470,0.7410}
\definecolor{coltr}{rgb}{0.8500,0.3250,0.0980}
\definecolor{coldm}{rgb}{0,0.4470,0.7410}
\definecolor{colbtr}{rgb}{0.9290,0.6940,0.1250}
\newcommand{\n}{{\bf n}}
\newcommand{\PP}{{\bf P}}
\newcommand{\Q}{{\bf Q}}
\begin{document}

\title{Surface-directed Dynamics in Living Liquid Crystals}  
\author{Aditya Vats}
\affiliation{Department of Physics, Indian Institute of Technology Delhi, New Delhi -- 110016, India.}`
\author{Varsha Banerjee}
\affiliation{Department of Physics, Indian Institute of Technology Delhi, New Delhi -- 110016, India.}
\author{Sanjay Puri}
\affiliation{School of Physical Sciences, Jawaharlal Nehru University, New Delhi -- 110067, India.}

\begin{abstract}
We study living liquid crystals (LLCs), which are an amalgam of nematic liquid crystals (LCs) and active matter (AM). These LLCs are placed in contact with surfaces which impose planar/homeotropic boundary conditions on the director field of the LC and the polarization field of the AM. The interplay of LC-AM interactions and the surface-directed conditions yield controlled pattern dynamics in the LLC, which has important technological implications. We discuss two representative examples of this pattern dynamics.
\end{abstract}

\maketitle

\section{Introduction}
\label{s1}

Active matter (AM) is an important example of an inherently nonequilibrium system which exhibits coherent dynamics on a much larger scale than the constitutent units. These units can be biological or synthetic, ranging from micrometres to meters, e.g., polar gels, bacterial suspensions, micro-tubule bundles, cytoplasmic streaming, bird flocks, fish shoals, vibrating granular rods, etc. \cite{Ramaswamy_2010, Marchetti_2013}. Active particles continuously consume energy from the surroundings and convert it to motion. The suspensions of particles like bacteria or synthetic swimmers have been the simplest and most characterized realizations of active systems. They reveal interesting features like reduction in effective viscosity, enhanced self-diffusion, and active clustering, distinctive from equilibrium colloidal suspension \cite{Marchetti_2013, Bechinger_2016}. AM also holds promise for creating miniature machines. For example, a bacterial bath can generate persistent motion in tiny gears utilized as devices for harvesting energy \cite{Angelani_2009, Sokolov_2010, Leonardo_2010}. The turbulent motion of AM has also been harnessed to enable immersed rotors to operate as micropumps \cite{Thampi_2016}. 

Although apparently coordinated, AM can exhibit irregular behavior on larger scales as collisions with the medium particles cause random tumblings \cite{Bechinger_2016}. The utility of AM can be greatly enhanced with the possibility of organized or tailored flows. For example, directed trajectories can be utilized in cargo transport or targeted drug delivery \cite{Kaiser_2014, Mathijssen_2018, Sokolov_2015}. A contemporary system of relevance in this context is {\it living liquid crystals} (LLCs), or a suspension of active particles such as bacteria in nematic liquid crystals (LCs) \cite{Zhou_2014, Peng_2016, Genkin_2017, Zhou_2018, Trivedi_2015, Aranson_2018, Aditya_2023}. In the ordered state, the rod-shaped LC molecules (nematogens) align along a preferred direction called the {\it director}, while maintaining positional fluidity \cite{Stephen_1974}. The strong LC-AM coupling substantially alters the collective behavior of the two-component system. An important experimental fact in LLCs is the co-alignment of the active particles and the nematogens \cite{Zhou_2014, Mushenheim_2014, Turiv_2020}. Consequently, the active particles swim parallel to the director, and topological defects in LCs play a significant role in transporting these swimmers. Experiments have reported a preferential movement of bacteria from defects with $-1/2$ charge towards defects with $+1/2$ charge \cite{Genkin_2017, Aranson_2018}. The active particles also perturb the director field at macroscopic length scales, and reveal important information about the visco-elastic properties of the LC medium. Further, the self-propulsion energy gets stored in director perturbations that can be harnessed into useful work \cite{Lapointe_2004, Musevic_2006}. Clearly, LLCs offer many pathways for the control of one component (AM or LC) by the other (LC or AM). This opens up the possibility of diverse scientific and technological applications.

Many experiments have shown that that boundary conditions imposed at container surfaces can significantly impact pattern formation and dynamics in LLCs \cite{Ravnik_2013,  Wensink_2008, Wioland_2013, Lushi_2014, Zhou_2017, Fielding_2011, Fily_2014}. The surface-directed dynamics differs substantially from the bulk dynamics, and often exhibits novel features. It has been reported that boundaries can act as sources or sinks of orientational order, leading to patterns such as stripes, vortices, and clusters \cite{Hernandez_2005, Kudrolli_2008, Lushi_2014, Yang_2014, Deseigne_2012}. They are also known to self-organize and stabilize patterns that would otherwise be unstable in the bulk. In an important experiment, Peng et al. \cite{Peng_2016} generated a pre-determined bulk configuration by appropriate surface treatment of the bounding plates.

The effect of confinement has been theoretically well-studied in the context of pure LCs \cite{ Wells_2007, Tsakonas_2007, Kusumaatmaja_2015, konark_2019, Konark_2019_2}. Boundary anchoring on nematic-filled square wells has been utilized to control the topological defects and obtain tailored morphologies. In particular, LC square wells are known to be bistable without any external field \cite{Barberi_1998, Tsakonas_2007, Kitson_2008}. It is natural to pose similar theoretical questions in the context of LLCs. For example, can tailored structures in LCs be exploited for directed transport of AM? Alternatively, can tailored flows dictated by boundary conditions on AM be used to create novel configurations in LCs? Finally, and perhaps most important, can the symbiotic interplay of these components plus boundary conditions yield hitherto unknown states of AM and LCs? The dual possibility of (a) tailoring active trajectories around novel defect configurations in LCs, and (b) the erasure of topological defects in LCs by active flows offers intriguing design concepts for microfluidic devices. We will address these and related questions from a theoretical perspective in the present paper.

In this work, we study the effect of several experimentally relevant boundary conditions imposed on director patterns and active flows. We use the phenomenological kinetic model for LLCs developed in our recent work \cite{Aditya_2023}. This model consists of the Toner-Tu (TT) model for AM, the time-dependent Ginzburg-Landau (TDGL) model for the LC, and an experimentally motivated coupling term which favors co-alignment of the two components. Our theoretical studies of this model in the bulk demonstrated two novel steady states: {\it chimeras} and {\it solitons}, which sweep through the coupled system in synchrony. Further, the symbiotic dynamics of the AM and LC can be exploited to induce and manipulate order in a component which is intrinsically (i.e., in the absence of coupling) disordered. In this work, we consider LLCs confined in square wells, and study the interplay of coupling and confinement on pattern dynamics. Some of the natural questions to address are as follows: Will the nematic morphologies remain stable in the presence of active entities? Can we control the active particle trajectories by tailoring the topological defects in the nematic component? How do different boundary conditions affect the emergent structures from this coupling between nematic order and activity?

This paper is organized as follows. In Sec.~\ref{s2}, we describe our coarse-grained modeling of LLCs. We also discuss typical boundary conditions which are imposed on the order parameters. In Sec.~\ref{s3}, we present detailed numerical results from our simulations of these models. We conclude this paper with a summary and discussion in Sec.~\ref{s4}.

\section{Modeling of Living Liquid Crystals}
\label{s2}

We employ an order-parameter-based description for the LLCs. This phenomenological approach uses free energy functionals based on the symmetries of the order parameters describing the system. For the nematic component, the order parameter is the ${\bf Q}$-tensor, which is a symmetric, traceless matrix whose leading eigenvector is the director ${\bf n}$. The components of ${\bf Q}$ can be written as $Q_{ij} = \mathcal{S}(n_i n_j - \delta_{ij}/d)$, where $d$ is the dimensionality. The scalar order parameter $\mathcal{S}$ measures the degree of orientational order about ${\bf n}$. For example, $\mathcal{S}=1$ describes a fully aligned nematic state, and a disordered state corresponds to low order with $\mathcal{S} \simeq 0$. The {\it Landau-de Gennes} (LdG) free energy for the LC can be written as \cite{Aditya_2020, Aditya_2021, Aditya_2022, Aditya_2023, konark_2019, Konark_2019_2, Stephen_1974}
\begin{eqnarray}
\label{LdG_FE}
F_Q[{\bf Q}] &=& \int \mbox{d}{\bf r} \left\{ \frac{A}{2}\mbox{Tr}(\boldsymbol{Q}^2)+\frac{B}{3}\mbox{Tr}(\boldsymbol{Q}^3)+\frac{C}{4}[\mbox{Tr}(\boldsymbol{Q}^2)]^2 + \frac{L}{2}\left|\nabla{\bf Q}\right|^2\right\}, 
\end{eqnarray}
where $A,\ B,\ C\ $ and $L$  are phenomenological parameters. We have $A=A_0(T-T_c)$,  where $A_0$ is a material-dependent coefficient and $T_c$ is the critical temperature for LC ordering. The gradient term in Eq.~\eqref{LdG_FE} penalises local variations in the order parameter.

Although AM is inherently nonequilibrium, one can formulate the TT model via a suitable ``free energy''. This is defined as a functional of the two ``order parameters": local density $\rho ({\bf r},t)$ (which is conserved) and the polarisation field ${\bf P}({\bf r},t)$ (which is nonconserved). The quantity ${\bf P}({\bf r},t)$ measures the local orientation of AM. The free energy in terms of these order parameters is given by \cite{Marchetti_2013}:
\begin{eqnarray}
F_a[\rho, {\bf P}] = \int \text{d}{\bf r} \left[\frac{\alpha(\rho)}{2}|{\bf P}|^2+\frac{\beta}{4}|{\bf P}|^4 +\frac{\kappa}{2}|\nabla {\bf P}|^2+ \frac{w}{2}|{\bf P}|^2\nabla\cdot {\bf P} -\frac{v_1}{2} (\nabla\cdot{\bf P}) \frac{\delta \rho}{\rho_0}+ \frac{D_\rho}{2}({\delta \rho})^2\right] , \label{fa}
\end{eqnarray}
where $\alpha, \beta, \kappa,  w, v_1, D_{\rho}$ are material-dependent parameters whose values can be related to the microscopic properties of the active particles \cite{Bertin_2009}. In Eq.~\eqref{fa}, $\delta \rho= \rho-\rho_0$, where $\rho_0$ is the average density of the system. The parameter $\alpha(\rho)=\alpha_0 (1-\rho/\rho_c)$, where $\alpha_0$ is a positive constant and  $\rho_c$ is the critical density. This free energy yields a continuous phase transition from a disordered state (with  $\rho=\rho_0$ and ${\bf P}=0$ for $\rho_0<\rho_c$) to an ordered state (with $\rho=\rho_0$ and $|{\bf P}| \sim \sqrt{(\rho_0/\rho_c-1)}$ for $\rho_0>\rho_c$). The TT model is written in terms of $F_a$ \cite{Marchetti_2013}, as will be discussed shortly. For the TT model with densities just above the transition point ($\rho \gtrsim \rho_0^+$), the ordered phase is unstable, and the system relaxes to a banded state that sweeps the system with speed $v_0$ (which is the same as the speed of active particles).

In recent work \cite{Aditya_2023}, we have proposed that the ``free energy'' of LLCs can be written as the sum of $F_Q$ and $F_a$, along with a suitable coupling term $F_c$: $F[{\bf Q}, \rho, {\bf P}] = F_Q[{\bf Q}]+F_a[\rho, {\bf P}]+F_c[{\bf Q}, \rho, {\bf P}]$. Experimental observations on LLCs dictate that active particles tend to align along the nematic director, i.e., $ \PP \parallel \n$ \cite{Zhou_2014, Genkin_2017}. The dyadic product of the \Q-tensor and the polarisation vector $\PP$ is the lowest order term that ensures this co-alignment and the ${\bf n} \rightarrow -{\bf n}$ symmetry. With these considerations, the free energy contribution from the coupling can be written as 
\begin{equation}
F_c[{\bf Q}, {\bf P}] =- c_0 \sum_{i,j} { Q_{ij}}P_i P_j,
\label{fe}
\end{equation}
where $c_0$ is the strength of the AM-LC interaction. The coupling term in terms of the director $\n$ is $-c_0({\bf n} \cdot {\bf P})^2$, which manifestly promotes co-alignment if $c_0 > 0$. For $c_0 < 0$, the preferred orientation is ${\bf P} \perp {\bf n}$. It is also possible to model the situation where ${\bf P}$ and ${\bf n}$ are aligned at an arbitrary angle, though the corresponding $F_c$ is more complicated than Eq.~(\ref{fe}).

The dissipative dynamics of LCs is studied via the TDGL equation for nonconserved kinetics \cite{Puri_2009}:
\begin{equation}
\frac{\partial {\bf Q}}{\partial t} = - \Gamma_Q \frac {\delta F_Q} {\delta {\bf Q}} ,
\end{equation}
where $\Gamma_Q$ is the kinetic coefficient. For simplicity, we present the corresponding LLC equations in $d=2$ here. In that case, ${\bf Q}$ is a $2\times 2$ matrix:
\begin{equation}
{\bf Q}=
\begin{pmatrix}
Q_{11} & Q_{12} \\
Q_{12} & -Q_{11}
\end{pmatrix},
\label{w}
\end{equation}
For the nematic components in LLCs, the corresponding TDGL equations are as follows \cite{Aditya_2023}:
\begin{eqnarray}
\frac{\partial {{Q_{11}}}}{\partial t} &=& -\Gamma_{{\bf Q}} \frac{\delta F_Q[\mathbf{Q}]}{\delta {{Q_{11}}}} + \Gamma_{{\bf Q}} c_0 (P_1^2-P_2^2),\label{tdgl_1}\\ 
\frac{\partial {{Q_{12}}}}{\partial t} &=& -\Gamma_{{\bf Q}} \frac{\delta F_Q[\mathbf{Q}]}{\delta {{Q_{12}}}} + \Gamma_{{\bf Q}} c_0 (2P_1P_2).
\label{tdgl_2}
\end{eqnarray}
Here, the damping parameter $\Gamma_Q$ sets the relaxation time scale for the system. The first term on the right-hand-side of Eqs.~\eqref{tdgl_1}-\eqref{tdgl_2} relaxes the pure LC to its free energy minimum. The second term represents the correction due to the coupling with AM. 

The AM dynamics is governed by the TT equations for the density and polarisation fields \cite{Toner_1995, Toner_1998}. As mentioned earlier, the TT model for pure AM can be formulated using a ``free energy" $F_a$ \cite{Marchetti_2013}. The incorporation of the coupling term $F_c$ in $F_a$ yields the relevant $d=2$ equations for AM in LLCs as follows \cite{Aditya_2023}:
\begin{eqnarray}
\frac{\partial \rho}{\partial t}&=& -v_0 \nabla \cdot({\bf P}\rho) - \nabla \cdot \left(-\Gamma_{\rho}\nabla \frac{\delta F_a}{\delta \rho}\right), \\\label{DD}
 \frac{\partial { P_1}}{\partial t}&=& \lambda_1 ({\bf P} \cdot \nabla ){P_1} - \Gamma_{{\bf P}}\frac{\delta F_a}{\delta { P_1}}  + \Gamma_{{\bf P}}c_0(Q_{11}P_1+Q_{12}P_2), \label{DP_1}\\
\frac{\partial { P_2}}{\partial t}&=& \lambda_1 ({\bf P} \cdot \nabla ){P_2} - \Gamma_{{\bf P}}\frac{\delta F_a}{\delta {P_2}}  + \Gamma_{{\bf P}}c_0 ( Q_{12}P_1-Q_{11}P_2). \label{DP_2}
\end{eqnarray}  

Here, $\Gamma_{\rho}$ and $\Gamma_{\PP}$ set the relaxation time scales for the density and polarisation fields. The first term on the RHS of Eqs.~\eqref{DP_1}-\eqref{DP_2} describes the effect of advection on the flow alignment -- $\lambda_1$ has the dimensions of speed. The terms with $c_0$ in Eqs.~\eqref{DP_1}-\eqref{DP_2} model the effect of the AM-LC coupling. The dimensionless versions of Eqs.~\eqref{tdgl_1}-\eqref{DP_2} are provided in Appendix \ref{appen_1}. We use these coupled equations as a model for LLCs in this letter. 

\begin{figure}[H]
\centering
\includegraphics[width=0.7\linewidth]{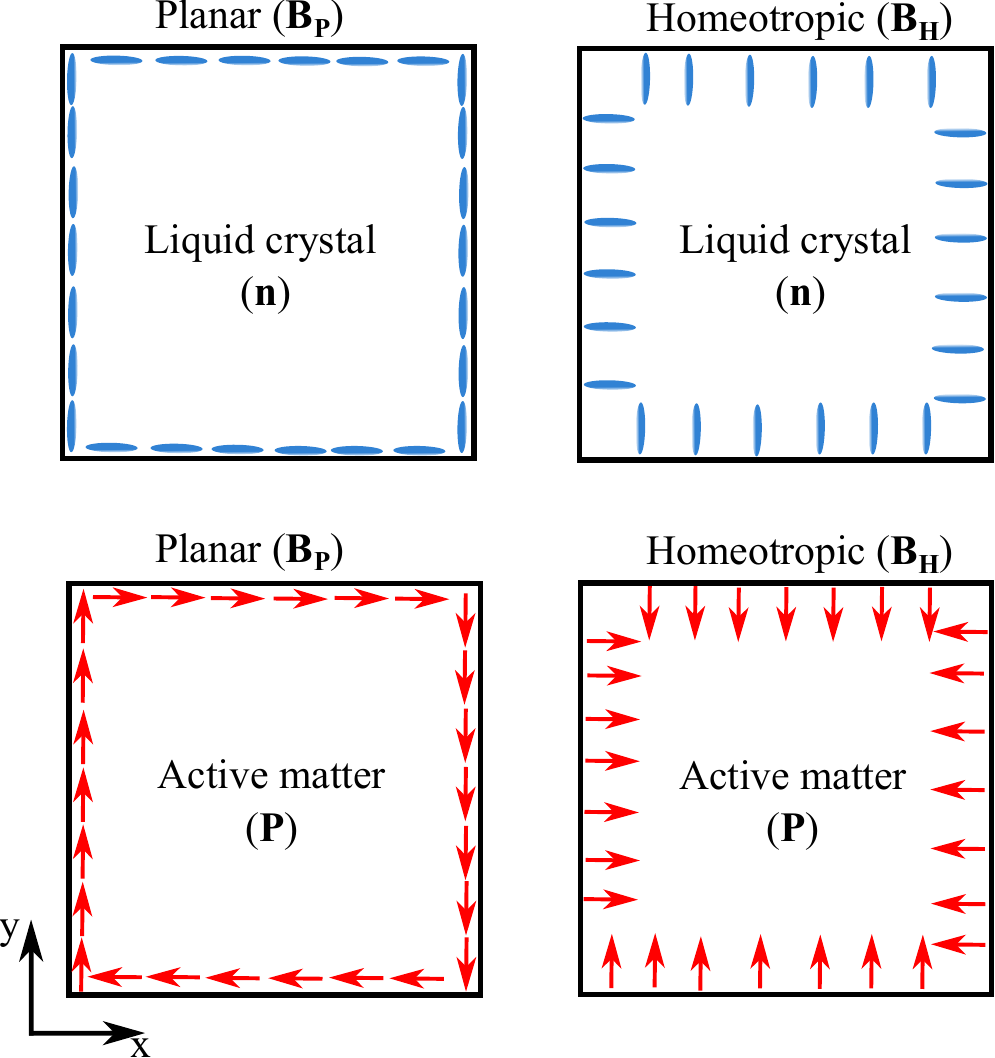} 
\caption{Schematic depicting planar ($B_P$) and homeotropic ($B_H$) boundary conditions for nematic (upper frames) and active (lower frames) components.}
\label{f1}
\end{figure}

In our previous work \cite{Aditya_2023}, we focused on symbiotic dynamics of this model in the bulk. Our focus in the present letter is the opposite, viz., whether we can control the dynamics of the AM and LCs by placing them in contact with a surface which exerts specific boundary conditions. We are motivated by the possibility of regulating pattern dynamics in the context of technological applications. Some quantitative statements about the nature of boundary conditions and the corresponding solutions in pure LCs and AM are in order. In the present work, we have studied all possible combinations of planar ($B_P$) and homeotropic ($B_H$) conditions for both components (see the schematic in Fig.~\ref{f1}). The vector order parameter (${\bf n}$ or ${\bf P}$) is anchored parallel (perpendicular) to the walls in planar (homeotropic) conditions. There are also additional possibilities for boundary conditions. These arise by (a) mixing $B_P$ and $B_H$ for ${\bf n}$ and ${\bf P}$ on different surfaces in Fig.~\ref{f1}; and (b) by flipping ${\bf P} \rightarrow -{\bf P}$ on ore or more surfaces in the lower frames of Fig.~\ref{f1}. For simplicity, we will restrict ourselves to the conditions depicted in Fig.~\ref{f1}.

These conditions are well-established in the literature for LC square wells \cite{Wells_2007, Tsakonas_2007, Kusumaatmaja_2015, konark_2019, Konark_2019_2}, and can be readily implemented in a physical setting. For example, the surface can be treated chemically to favor specific anchoring conditions on the nematic director. Other mechanisms to achieve the desired director orientation for LCs include lithography, surface anchoring, flow alignment and coupling to an external field \cite{Alexander_2012, Stephen_1974}. In the context of AM,  planar boundary conditions are most commonly used. The presence of geometric constraints is enough to implement these in experiments. For homeotropic conditions, experiments suggest that the confining walls can comprise different particles, and the potential between the active and wall particles can be tuned to achieve the desired boundaries \cite{Lowen_2008}. Setting up physical barriers or optical traps, pre-defined surface patterning and chemical modification are other commonly used techniques employed to set boundary conditions in AM \cite{Bechinger_2016}.

\section{Numerical Results}
\label{s3}

We have used the Euler discretization method \cite{kincaid_2009} to numerically solve Eqs.~\eqref{eq_Q1}-\eqref{Rho} and determine the evolution of the nematic and active components. The discretization mesh sizes used in our simulations are $\Delta x=\Delta y=1$ and $\Delta t=0.01$. The initial fields ${\bf Q}({\bf r},0)$ and ${\bf P}({\bf r}, 0)$ consist of small random fluctuations about 0, corresponding to the disordered state. Similarly, the initial density field $\rho ({\bf r},0)$ consists of small fluctuations about $\rho_0$. Thus, we study coarsening of the system from a disordered state at $t=0$. Notice that all the coupling terms in Eqs.~\eqref{eq_Q1}-\eqref{eq_P22} are quadratic. Therefore, the growth of linear fluctuations about the disordered state is the same as for the uncoupled system ($c_0=0$). The effect of the coupling is manifested only when the fields enter a nonlinear growth regime. The equations are solved using the different boundary conditions in Fig.~\ref{f1} for ${\bf n}$ and ${\bf P}$. We use periodic boundary conditions for $\rho$. The parameters are such that $T<T_c$ ($+$ sign in Eqs.~\eqref{eq_Q1}-\eqref{eq_Q2}) and $\rho=\rho_c^+=0.52$. In the bulk (mimicked by periodic boundary conditions for all fields), these parameters yield a uniform ordered phase for the nematic, and a banded state for the AM \cite{Aditya_2023}.
 
\begin{figure}[H]
\centering

\includegraphics[width=0.8\linewidth]{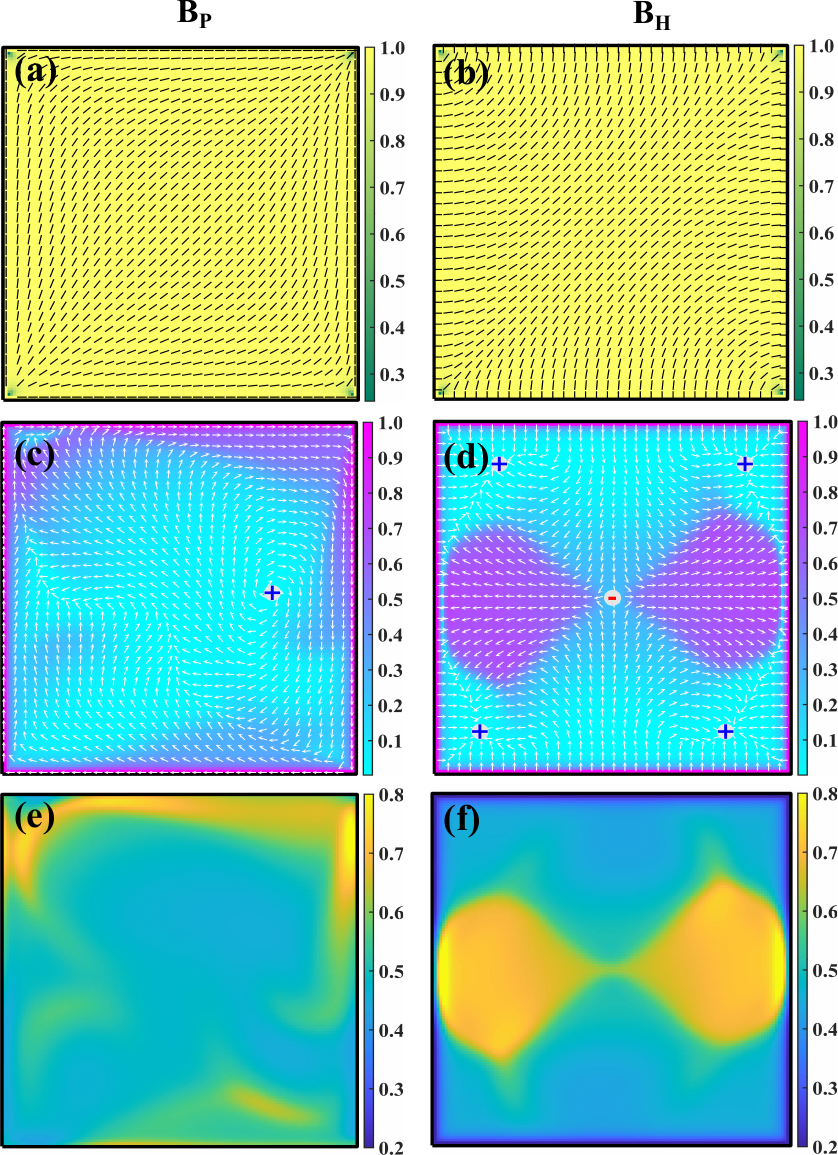} 
\caption{Snapshots at $t=10^4$ for the ${\bf n}$-field (first row), ${\bf P}$-field (second row), and $\rho$-field (third row) for $c_0=0$. The frames (a), (c) and (e) correspond to planar ($B_P$) boundary conditions. The color bars in these frames show the nematic orientational order $\mathcal{S}$ in (a); magnitude of polarisation $|{\bf P}|$ in (c); and density $\rho$ in (e). The rods (arrows) denote the orientation of the director (polarisation) field. The defects are denoted by $+$ or $-$, according to their signs. The corresponding snapshots for homeotropic ($B_H$) boundaries are shown in frames (b), (d) and (f).}
\label{f2}
\end{figure}

We start by discussing the consequences of planar and homeotropic boundary conditions for uncoupled systems, i.e., $c_0=0$. This will provide a reference point to judge the effect of the coupling. There have been many studies of both nematic \cite{konark_2019, Konark_2019_2, Wells_2007, Tsakonas_2007, Walton_2018, Kusumaatmaja_2015} and active components \cite{Ravnik_2013, Green_2017, Wensink_2008, Wioland_2013, Wu_2017, Lushi_2014, Zhou_2017, Fielding_2011, Fily_2014, Hernandez_2005, Kudrolli_2008, Yang_2014, Deseigne_2012} in this context. In this limit, both fields evolve independently. Figs.~\ref{f2} (a)-(b) show nematic morphologies at $t=10^4$ for planar ($B_P$) and homeotropic ($B_H$) boundary conditions. The color bar indicates the magnitude of the orientational order parameter $\mathcal{S}$. A topological defect of charge $+1/2$ ($-1/2$) in the nematic medium is identified as a point around which the orientation of the (apolar) director changes by $+\pi$ ($-\pi$) when traversed clockwise. Naturally, the order parameter $\mathcal{S} \simeq 0$ at the defect. In Figs.~\ref{f2} (a)-(b), the director ${\bf n}$ (denoted by rods) aligns diagonally in the square well indicating the absence of bulk defects for both $B_P$ and $B_H$. There are topological structures reminiscent of partial defects at the corners and surfaces, but these do not categorize as full defects and will not be discussed further. The corresponding ${\bf P}$-field for the active component is shown in Figs.~\ref{f2}(c) and \ref{f2}(d). The arrows represent the orientation {\bf P}. The active medium has $+1$ ($-1$) defects, corresponding to points where ${\bf P}$ rotates by $2\pi$ ($-2\pi$) when traversed clockwise. For $B_P$, there is a single $+1$ defect which moves around in the system. For $B_H$, there are multiple defects in the system [as shown in Fig.~\ref{f2}(d)] with a complicated dynamical interplay. The supplementary material (SM) shows movies (\href{https://csciitd-my.sharepoint.com/:v:/g/personal/phz178374_iitd_ac_in/EXLCbNr7IpxDrDDWZcE5RUEB93SR-x54JeO53Yz1Gw5p6A?nav=eyJyZWZlcnJhbEluZm8iOnsicmVmZXJyYWxBcHAiOiJPbmVEcml2ZUZvckJ1c2luZXNzIiwicmVmZXJyYWxBcHBQbGF0Zm9ybSI6IldlYiIsInJlZmVycmFsTW9kZSI6InZpZXciLCJyZWZlcnJhbFZpZXciOiJNeUZpbGVzTGlua0NvcHkifX0&e=mdmbmG}{M1}) of the evolution for Figs.~\ref{f2}(c)-(d). The color bar adjacent to Figs.~\ref{f2}(c)-(d) denotes the magnitude $|{\bf P}|$ -- this goes to 0 at the defect cores. As expected, $\rho$ in Figs.~\ref{f2}(e)-(f) tracks the ${\bf P}$-variation due to the $\rho$-${\bf P}$ coupling in the TT equations. We should stress that the morphologies in Figs.~\ref{f2} (a)-(b) are static, i.e., the various fields have settled to fixed-point values. However, the ${\bf P}$ field in Figs.~\ref{f2}(c)-(d) is dynamic.

Next, let us study the effect of the coupling on the surface-directed dynamics. We have examined all possible combinations of boundary conditions in this context. Here, we only show some representative results.

\begin{figure}[H]
\centering  
\includegraphics[width=0.9\linewidth]{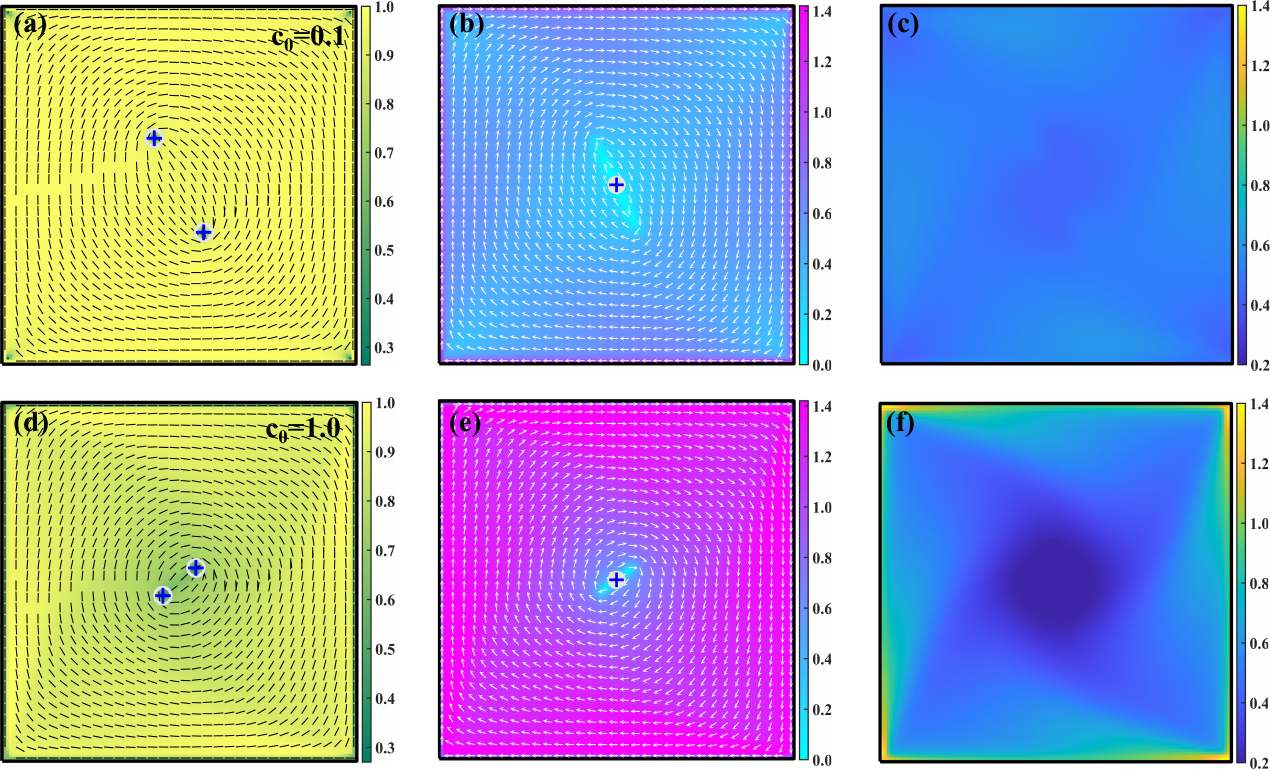}
\caption{Snapshots at $t=10^4$ for the coupled case with $c_0=0.1$ (upper row) and $c_0=1.0$ (lower row). $B_P$ boundary conditions are imposed at the surfaces for both ${\bf n}$ and ${\bf P}$. The frames (a), (d) show the ${\bf n}$-field; (b), (e) show the ${\bf P}$-field; and (c), (f) show the $\rho$-field. The meaning of various symbols and color bars is the same as in Fig.~\ref{f2}.}
\label{f3}
\end{figure}

First, we consider the coupled system with $B_P$ surfaces. Fig.~\ref{f3} presents the snapshots (at $t=10^4$) of the nematic and active components for $c_0=0.1$ (upper row) and 1.0 (lower row). In Figs.~\ref{f3}(a) and (d), we show the $\mathcal{S}$-field (see color bar) along with ${\bf n}$ (denoted by rods). The corresponding ${\bf P}$-field and its magnitude are depicted in Figs.~\ref{f3}(b) and (e). Notice the co-alignment of ${\bf n}$ and ${\bf P}$ due to the coupling. More importantly, the LC morphologies are no longer static as in Fig.~\ref{f2}(a). Here, the asymptotic LC state consists of two co-rotating topological defects (at a fixed distance $d_0$) with the same charge ($+1/2$) at the centre of the square well. These defects move closer with increasing $c_0$. The corresponding ${\bf P}$-field exhibits a vortex with $+1$ charge at the centre of the square well. The movies of the evolution for both coupling strengths can be found in \href{https://csciitd-my.sharepoint.com/:v:/g/personal/phz178374_iitd_ac_in/EUa6PbjAtutBrc166fpsmPkBoDaaIDrTCNDYHvO5LSrBAA?nav=eyJyZWZlcnJhbEluZm8iOnsicmVmZXJyYWxBcHAiOiJPbmVEcml2ZUZvckJ1c2luZXNzIiwicmVmZXJyYWxBcHBQbGF0Zm9ybSI6IldlYiIsInJlZmVycmFsTW9kZSI6InZpZXciLCJyZWZlcnJhbFZpZXciOiJNeUZpbGVzTGlua0NvcHkifX0&e=ls3EWj}{M2} ($c_0=0.1$) and \href{https://csciitd-my.sharepoint.com/:v:/g/personal/phz178374_iitd_ac_in/ERDfFJHmlEJHjWr4Dz7uJ_kBI2y2U3wxPwTegSC7-75uNg?nav=eyJyZWZlcnJhbEluZm8iOnsicmVmZXJyYWxBcHAiOiJPbmVEcml2ZUZvckJ1c2luZXNzIiwicmVmZXJyYWxBcHBQbGF0Zm9ybSI6IldlYiIsInJlZmVycmFsTW9kZSI6InZpZXciLCJyZWZlcnJhbFZpZXciOiJNeUZpbGVzTGlua0NvcHkifX0&e=SBghSB}{M3} ($c_0=1.0$) in the SM. In Figs.~\ref{f3}(c) and (f), we show the density field in the system. There is a large variation of $\rho$ in the strongly coupled ($c_0=1.0$) system -- the dilute regions coincide with the vortex core with $|{\bf P}| \simeq 0$. From an application perspective, such morphologies can be harnessed to create a pumping effect in microfluidic devices \cite{Thampi_2016}. The systematic circular motion generates fluid flow inside the devices, which can be channeled in any desired direction.

\begin{figure}[H]
\centering  
\includegraphics[width=0.9\linewidth]{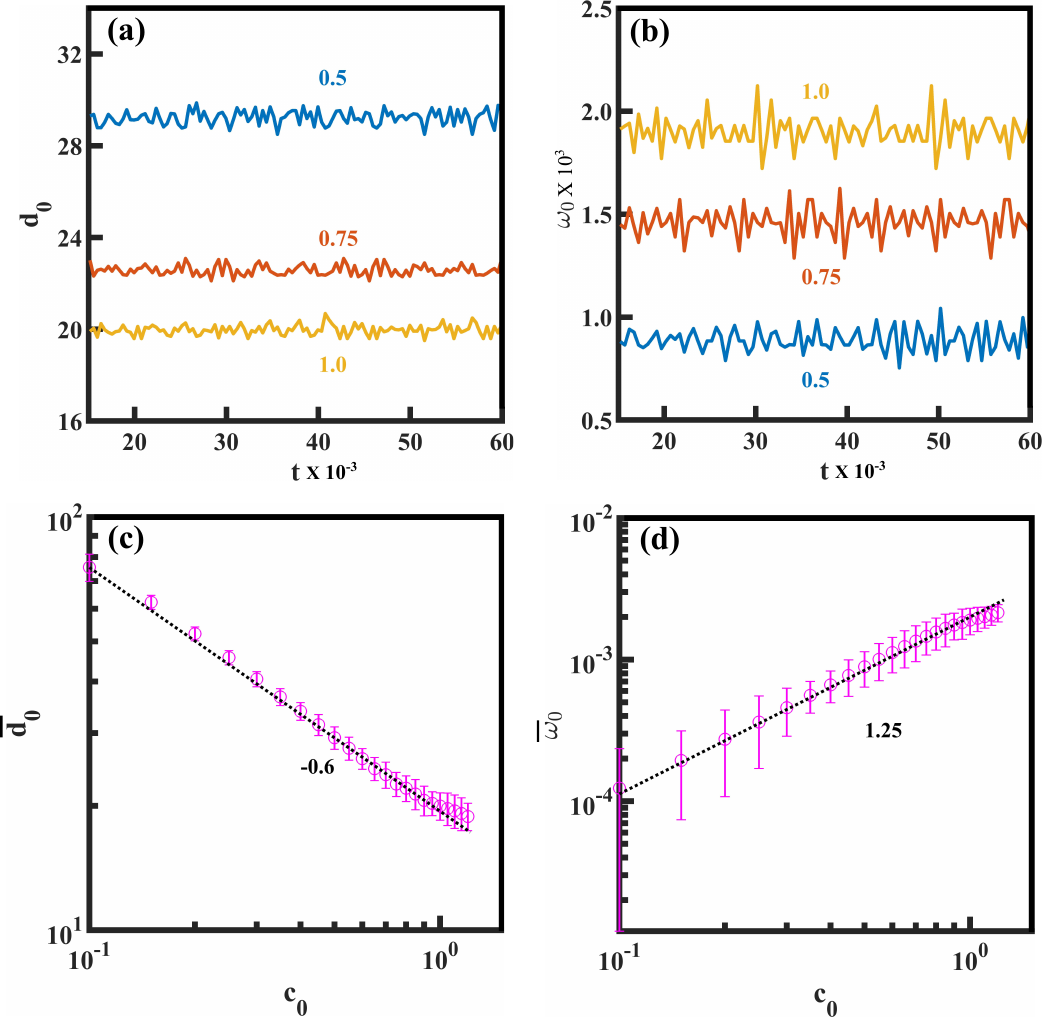}
\caption{(a) Plot of the inter-vortex distance $d_0$ vs. $t$ in the asymptotic state for $c_0=0.5,\ 0.75,\ 1.0$. (b) Plot of the co-rotation angular velocity $\omega_0$ vs. $t$ for the same $c_0$-values. (c) Log-log plot of $\bar{d}_0$ vs. $c_0$. The bar denotes the time-average in the asymptotic state. The dashed line denotes the best linear fit to the data. (d) Log-log plot of $\bar{\omega}_0$ vs. $c_0$.}
\label{f4}
\end{figure}

Before proceeding, we wish to quantify how the morphologies in Fig.~\ref{f3} change with $c_0$. The co-rotating vortices in the ${\bf n}$-field become more tightly bound as $c_0$ increases. In Fig.~\ref{f4}(a), we plot the inter-vortex distance $d_0$ vs. $t$ in the asymptotic state for $c_0 = 0.5, 0.75, 1.0$. The time-series fluctuates chaotically about an average value. In this context, we make two remarks. First, as the spatial mesh size is $\Delta x=1$, there are inaccuracies in determining the precise locations of the vortex cores. These become more marked at higher $c_0$ as the vortices come closer together. Second, in the square lattice, there is an intrinsic anisotropy depending on the relative alignment of the line connecting the vortex cores and the diagonal of the square well. We attribute the chaotic fluctuations in $d_0(t)$ vs. $t$ to these numerical factors. In Fig.~\ref{f4}(b), we plot the co-rotation angular velocity $\omega_0$ vs. $t$ in the asymptotic state for the same values of $c_0$. In Fig.~\ref{f4}(c), we plot $\bar{d_0}$ vs. $c_0$ on a log-log scale. Here, $\bar{d_0}$ represents the time average of $d_0(t)$ in the asymptotic state. We expect $\bar{d_0}\rightarrow \infty$ as $c_0 \rightarrow 0$, corresponding to the uncoupled limit. Our numerical data is consistent with a power-law behavior $\bar{d_0} \sim c_0^{-\theta}$ with $\theta \simeq 0.60$, though there is a saturation for $c_0 > 1$. In Fig.~\ref{f4}(d), we plot $\bar{\omega}_0$ vs. $c_0$ on a log-log scale. In the uncoupled limit ($c_0 \rightarrow 0$), we expect $\bar{\omega}_0 \rightarrow 0$. Our numerical data is again consistent with a power-law behavior $\bar{\omega}_0 \sim c_0^\alpha$ with $\alpha \simeq 1.25$ for $c_0 < 1$. What consequences do these observations have on AM? We get a flavor from Fig.~\ref{f3} -- the swirling is stronger with increasing $c_0$, and the AM is pushed closer to the periphery of the well. This is due to the interplay of the inherent linear velocity $v_0$ and the coupling-induced angular velocity $\omega_0$. These prototypical observations not only demonstrate the symbiotic relationship between LCs and AM, but also provide a systematic procedure for manipulating pattern formation via the coupling strength.

\begin{figure}[H]
\centering  
\includegraphics[width=0.9\linewidth]{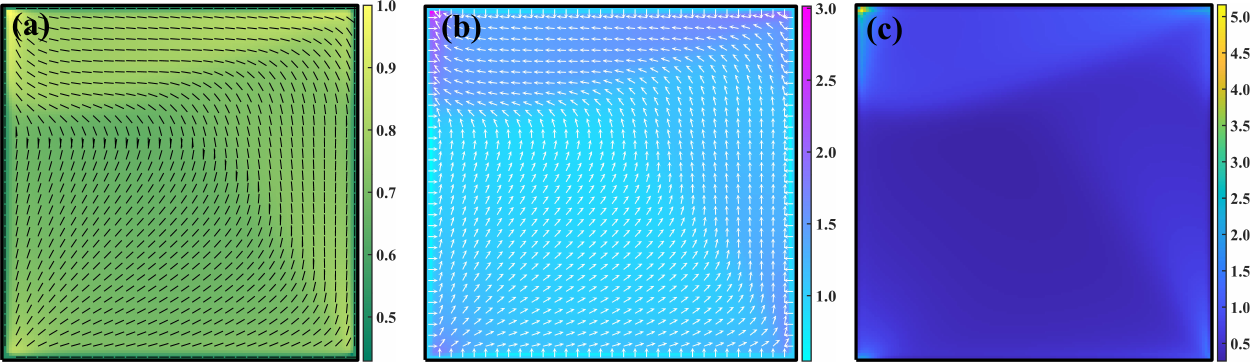}
\caption{Snapshots at $t=10^4$ for the coupled case with $c_0=1.0$. The boundary conditions for LCs and AM are $B_P$ and $B_H$, respectively. The frames show the (a) ${\bf n}$-field. (b) ${\bf P}$-field. (c) $\rho$-field. The color bars denote the magnitude of the relevant field.}
\label{f5}
\end{figure}

 Our second example of coupled kinetics is the case where the LC and AM have $B_P$ and $B_H$ boundary conditions, respectively. The resultant morphologies from our coarsening experiments for $c_0=1.0$ are shown in Fig.~\ref{f5}. The snapshots are shown at $t=10^4$, by which time the dynamis has settled to a 
fixed point (FP). In Fig.~\ref{f5}(a), we show the $\mathcal{S}$-field with director orientations for the nematogens. No defects are seen in the nematic field. The corresponding ${\bf P}$-field and its magnitude are shown in Fig.~\ref{f5}(b). The $\rho$-field is depicted in Fig.~\ref{f5}(c). In the uncoupled limit, the relevant configurations are shown in Figs.~\ref{f2}(a),(d), (f). Fig.~\ref{f2}(a) shows FP behavior, whereas Figs.~\ref{f2}(d), (f) show complex dynamical states with multiple defects. In Fig.~\ref{f5}, the coupling controls the dynamics of AM and harnesses it to an FP behavior.

\section{Summary and Discussion}
\label{s4}

Let us conclude this letter with a summary and discussion of our results. We have focused on the effect of {\it boundary conditions} (BCs) on the dynamics of {\it living liquid crystals} (LLCs). Our purpose is to examine whether surfaces can be tailored to inject specific dynamical behaviors into an LLC. This control is expected to yield a range of possible applications in science and technology. We consider two types of BC: (a) {\it planar} or $B_P$, where ${\bf n}$ or ${\bf P}$ are aligned parallel to the surfaces; and (b) {\it homeotropic} or $B_H$, where ${\bf n}$ or ${\bf P}$ point perpendicular to the surface. As ${\bf P} \neq -{\bf P}$, there are further sub-classes in $B_P$ and $B_H$ depending on the direction of {\bf P}. These BCs can arise naturally due to confinement of the LLC in a container. Alternatively, specific BCs may be imposed at surfaces to control the dynamics of the LLC. In this letter, we have shown two representative examples of LLCs in square wells. \\
(a) First, we consider the case where {\bf n} has $B_P$, and {\bf P} has $B_P$ with the directionality being cyclic along the surfaces. In the uncoupled limit, the ${\bf n}$-field is static, whereas the ${\bf P}$-field has a single vortex wandering in the system. In the coupled case, the system settles into a controlled dynamics with a co-rotating pair of defects in the ${\bf n}$-field. The spacing and angular velocity of this co-rotation have a power-law dependence on the coupling strength. \\
(b) Second, we consider the case where ${\bf n}$ and ${\bf P}$ have $B_H$, with ${\bf P}$ pointing inward from the surfaces. In the uncoupled limit, the ${\bf n}$-field is static, and the ${\bf P}$-field has a complex dynamics with multiple defects, swirling around. In the coupled case, this complex dynamics is tamed and both ${\bf n}$ and ${\bf P}$ settle into a steady state.

Generally speaking, AM is ubiquitous in nature. The constituent particles tend to parallelize locally, but can exhibit complex unstructured dynamics at the macroscopic level, e.g., turbulent motion. A major research direction in AM has focused on disciplining and harnessing their motion into useful work. Consequently, LLCs are emerging as valuable microfluidic devices with potential applications in sorting and mixing of materials, bio-sensing, and targeted drug delivery in bio-medical applications \cite{Zhang_2021}. We have demonstrated in this letter that BCs play a crucial role in pattern dynamics in LLCs. An improved understanding of the interplay between LLCs and surfaces can help design active systems with specific pattern dynamics. In this letter, we have studied defect dynamics in LLCs, and investigated the interplay of AM-LC interactions and confining surfaces. In this context, we displayed two important examples from a plethora of dynamical possibilities, e.g., (a) harnessing of random motion into a controlled dynamical trajectory; and (b) taming of a dynamical state to a static state. These examples provide a flavor of the possibilities of surface-directed dynamics in LLCs. We believe that surface-directed behavior opens up the possibility of several novel applications, e.g., active morphologies with persistent motion around a defect core can be used as microfluidic pumps. We hope our present theoretical study will guide future experiments on LLCs, and pave the way for their utilization in devices.

\subsection*{Acknowledgements}

AV acknowledges UGC, India for support via a research fellowship. VB acknowledges DST, India for research grants. AV and VB gratefully acknowledge the HPC facility of IIT Delhi for computational resources.

\newpage
\appendix

\section{Dynamical Equations for Living Liquid Crystals}
\label{appen_1}

The dynamical Eqs.~\eqref{tdgl_1}-\eqref{DP_2} for $d=2$ living liquid crystals (LLCs) are presented in an expanded form in this appendix. It is easier to work with the dimensionless form due to the reduced number of parameters, and the identification of universal spatial and temporal scales. For the sake of brevity, we only present the dimensionless forms of Eqs.~\eqref{tdgl_1}-\eqref{DP_2}. These are obtained by introducing rescaled variables as
\begin{eqnarray}
&& {\bf Q} = c_Q{\bf Q}^\prime, \quad {\bf P} = c_P{\bf P}^\prime, \quad {\bf r} =c_r {\bf r}^\prime, \quad t= c_t t^\prime, \quad \mbox{where} \nonumber \\
&& c_Q = \sqrt{\frac{|A|}{2C}}; \quad c_P=\sqrt{\frac{\alpha_0}{\beta}}; \quad c_t=\frac{\beta}{\alpha_0\Gamma_Q} \sqrt{\frac{|A|}{2C}}; \quad
c_r=\sqrt{\frac{L}{|A|}} .
\end{eqnarray}
Dropping the primes, the dimensionless equations can be written as
\begin{eqnarray}
\frac{\partial Q_{11}}{\partial t}&=&\xi_1\left[\pm Q_{11}-(Q_{11}^2+Q_{12}^2)Q_{11}+\nabla^2Q_{11}\right]+c_0 (P_1^2-P_2^2), \label{eq_Q1} \\
\frac{\partial Q_{12}}{\partial t}&=&\xi_1 \left[\pm Q_{12}-(Q_{11}^2+Q_{12}^2)Q_{12}+\nabla^2Q_{12}\right]+ 2c_0 P_1P_2,\label{eq_Q2} \\
\frac{1}{\Gamma}\frac{\partial P_1}{\partial t}&=& \xi_2 \bigg[\left(\frac{\rho}{\rho_c}-1-{\bf P}\cdot {\bf P}\right) P_1 - \frac{v_1^\prime}{2\rho_0} \nabla_x \rho + \lambda_1^\prime ({\bf P} \cdot \nabla)P_1 +\lambda_2^\prime \nabla_x (|{\bf P}|^2) \nonumber \\
&& + \lambda_3^\prime P_1(\nabla \cdot {\bf P}) + \kappa^\prime \nabla^2 P_1 \bigg] + c_0 (Q_{11}P_1+Q_{12}P_2), \label{eq_P11} \\
\frac{1}{\Gamma}\frac{\partial P_2}{\partial t}&=&\xi_2 \bigg[\left(\frac{\rho}{\rho_c}-1-{\bf P}\cdot {\bf P}\right) P_2 - \frac{v_1^\prime}{2\rho_0} \nabla_y \rho + \lambda_1^\prime({\bf P} \cdot \nabla)P_2 +\lambda_2^\prime \nabla_y (|{\bf P}|^2) \nonumber \\
&& +\lambda_3^\prime  P_2(\nabla \cdot {\bf P}) + \kappa^\prime\nabla^2 P_2\bigg] + c_0 ( Q_{12}P_1-Q_{11}P_2), \label{eq_P22} \\
\frac{1}{\Gamma^\prime}\frac{\partial \rho}{\partial t}&=& -v_0^\prime \nabla\cdot({\bf P}\rho) + D_{\rho}^\prime \nabla^2 \rho. \label{Rho}
\end{eqnarray}

The $\pm$ sign in Eqs.~\eqref{eq_Q1}-\eqref{eq_Q2} determines whether the nematic component is above ($-$) or below ($+$) its critical temperature $T_c$. For $T < T_c$, the nematic is intrinsically (i.e., for $c_0=0$) ordered. For $T>T_c$, the nematic is intrinsically ordered. Similarly, the ${\bf P}$-field in Eqs.~\eqref{eq_P11}-\eqref{eq_P22} is intrinsically disordered if $\rho_0<\rho_c$, and intrinsically ordered for $\rho_0>\rho_c$.  The re-scaled parameters and their numerical values in our simulations are provided in Table~\ref{Tab1}. These parameter values are similar to those chosen for the uncoupled system (AM or LC) in the literature. However, we emphasize that our simulation results do not change qualitatively on changing the above values as long as the solutions are stable.
\begin{table}[H]
\begin{center}
      \begin{tabular}{|c|c|}
       \hline
         {\it Scaled Parameters} & {\it Numerical Values}
        \\ 
       \hline
      
        $\xi_1 = \dfrac{2|A|\beta}{\alpha_0}\sqrt{\dfrac{|A|}{2C}}, \quad
\xi_2=\dfrac{\alpha_0}{2} \sqrt{\dfrac{2C}{|A|}}$& 1,\ 1 \\
        \hline
        $v_1^\prime=\dfrac{v_1}{\alpha_0}\sqrt{\dfrac{\beta|A|}{\alpha_0L}}, \quad v_0^\prime = \dfrac{v_0}{\Gamma_{\rho}}\sqrt{\dfrac{\alpha_0|A|}{\beta L}}$& 0.5,\ 0.25 \\
        \hline
       
        $\Gamma=\dfrac{\beta|A|\Gamma_P}{\alpha_0\Gamma_QC}, \quad
\Gamma^\prime=\dfrac{\beta\Gamma_{\rho}}{\alpha_0\Gamma_Q}\sqrt{\dfrac{|A|}{2C}}$& 1,\ 1 \\

\hline

$ \kappa^\prime=\dfrac{\kappa |A|}{\alpha_0 L}, \quad
D_{\rho}^\prime=\dfrac{D_{\rho} |A|}{L}$ & 1,\ 1 \\

\hline
$\lambda_1^\prime=\dfrac{\lambda_1}{\Gamma_P}\sqrt{\dfrac{|A|}{\alpha_0\beta L}},  \quad \lambda_2^\prime= \lambda_2 \sqrt{\dfrac{|A|}{\alpha_0\beta L}}, \quad 
\lambda_3^\prime= \lambda_3 \sqrt{\dfrac{|A|}{\alpha_0\beta L}}$& $-0.5$,\ $-0.5$,\ $0.5$ \\
\hline

    \end{tabular}
      \end{center}
    \caption{Dimensionless parameters in Eqs.~\eqref{eq_Q1}-\eqref{Rho}, and their numerical values.} 
    \label{Tab1}
\end{table}

\bibliography{ref.bib}
\bibliographystyle{apsrev4-1}

\newpage

\end{document}